\documentclass[twocolumn,A4]{article}
\usepackage[dvips]{graphics}

\usepackage{setspace}
\usepackage{color}
\definecolor{gold}{rgb}{0.85,0.66,0}
\definecolor{dblue}{rgb}{0,0,0.8}

\begin{document}

\onecolumn

\begin{center}
{\bf{\Large {\textcolor{gold}{Electron transport through a quantum 
interferometer: A theoretical study}}}}\\
~\\
{\textcolor{dblue}{Santanu K. Maiti}}$^{\dag,\ddag,}$\footnote{{\bf 
Corresponding Author}: Santanu K. Maiti \\
$~$\hspace {0.45cm} Electronic mail: santanu.maiti@saha.ac.in} \\
~\\
{\em $^{\dag}$Theoretical Condensed Matter Physics Division,
Saha Institute of Nuclear Physics, \\
1/AF, Bidhannagar, Kolkata-700 064, India \\
$^{\ddag}$Department of Physics, Narasinha Dutt College,
129 Belilious Road, Howrah-711 101, India} \\
~\\
{\bf Abstract}
\end{center}
In the present work we explore electron transport properties through 
a quantum interferometer attached symmetrically to two one-dimensional
semi-infinite metallic electrodes, namely, source and drain. The 
interferometer is made up of two sub-rings where individual sub-rings
are penetrated by Aharonov-Bohm fluxes $\phi_1$ and $\phi_2$, respectively.
We adopt a simple tight-binding framework to describe the model and all 
the calculations are done based on the single particle Green's function 
formalism. Our exact numerical calculations describe two-terminal
conductance and current as functions of interferometer-to-electrode 
coupling strength, magnetic fluxes threaded by left and right 
sub-rings of the interferometer and the difference of these two fluxes. 
Our theoretical results provide several interesting features of electron 
transport across the interferometer, and these aspects may be utilized to
study electron transport in Aharonov-Bohm geometries.

\vskip 1cm
\begin{flushleft}
\end{flushleft}

\newpage
\twocolumn

\section{Introduction}

The study of electronic transport in quantum confined model systems like 
quantum rings, quantum dots, arrays of quantum dots, quantum dots embedded 
in a quantum ring, etc., has become one of the most fascinating branch of
nanoscience and technology. With the aid of present nanotechnological
progress~\cite{nanofab1,nanofab2}, these simple looking quantum confined 
systems can be used to design nanodevices especially in electronic as 
well as spintronic engineering. The key idea of 
manufacturing nanodevices is based on the concept of quantum 
interference effect~\cite{imry1,imry2,gefen1}, and it is generally 
preserved throughout the sample having dimension smaller or comparable 
to the phase coherence length. Therefore, ring type conductors or two 
path devices are ideal candidates to exploit the effect of quantum 
interference~\cite{bohm}. In a ring shaped geometry, quantum 
interference effect can be controlled by several ways, and most 
probably, the effect can be regulated significantly by tuning the 
magnetic flux, the so-called Aharonov-Bohm (AB) flux~\cite{chang,
han,he,wel}, that threads the ring. This key feature motivates us to 
widely use quantum interference devices in mesoscopic solid-state 
electronic circuits~\cite{roh}. Using a mesoscopic ring we can 
construct a quantum interferometer, and here we will show that the 
interferometer 
exhibits several exotic features of electron transport which can be 
utilized in designing nanoelectronic circuits. To reveal the phenomena, 
we make a bridge system, by inserting the interferometer between two 
electrodes (source and drain), the so-called source-interferometer-drain 
bridge. Following the pioneering work of Aviram and Ratner~\cite{aviram},
theoretical description of electron transport in a bridge system has got 
much progress. Later, many excellent experiments~\cite{holl,
kob,kob1,ji,yac,reed1,reed2} have been done in several bridge systems 
to understand the basic mechanisms underlying the electron transport. 
Though extensive studies on electron transport have already been done 
both theoretically~\cite{chang,han,he,wel,orella1,orella2,nitzan1,
nitzan2,bai,muj1,muj2,cui,baer2,baer3,tagami,walc1,baer1,maiti1,
maiti2,maiti3,walc2,gefen,kubo,naka,aharony,konig,ern2} as well as 
experimentally~\cite{holl,kob,kob1,ji,yac,reed1,reed2}, yet lot of 
controversies are still present between the theory and experiment, and 
the complete description of the conduction mechanism in this scale is 
not very well defined even today. Several controlling 
factors are there which can significantly regulate electron transport in 
a conducting bridge, and all these factors have to be taken into account 
properly to understand the transport mechanism. For our illustrative 
purposes, here we mention some of these issues.
\begin{enumerate}
\item The quantum interference effect~\cite{baer2,baer3,tagami,walc1,
baer1,maiti1,maiti2} of electronic waves passing through different 
arms of the bridging material becomes the most significant issue. 
\item The coupling of the bridging material to the electrodes significantly
controls the current amplitude across any bridge system~\cite{baer1}. 
\item Dynamical fluctuation in small-scale devices is another important 
factor which plays an active role and can be manifested through the 
measurement of {\em shot noise}~\cite{walc2}, a direct consequence of 
the quantization of charge.
\item Geometry of the conducting material between the two electrodes
itself is an important issue to control electron transmission which
has been described quite elaborately by Ernzerhof 
{\em et al.}~\cite{ern2} through some model calculations. 
\end{enumerate}

Addition to these, several other factors of the tight-binding 
Hamiltonian that describe a system also provide important effects in 
the determination of current across a bridge system.

In this presentation we explore electron transport properties of a 
quantum interferometer based on the single particle Green's function 
formalism. The interferometer is sandwiched between two semi-infinite
one-dimensional ($1$D) metallic electrodes, viz, source and drain, and, 
two sub-rings of the interferometer are subject to the Aharonov-Bohm (AB) 
fluxes $\phi_1$ and $\phi_2$, respectively. The schematic view of the
bridge system is depicted in Fig.~\ref{ring}. A simple tight-binding model 
is used to describe the system and all the calculations are done 
numerically, which illustrate conductance-energy and current-voltage 
characteristics as functions of the interferometer-to-electrode coupling 
strength, magnetic fluxes and the difference of these two fluxes. 
Several exotic features are observed from this study. These are: (i) 
semiconducting or metallic nature depending on the coupling strength of 
the interferometer to the side attached electrodes, (ii) appearance of 
anti-resonant states~\cite{chang,han,he} and (iii) unconventional 
periodic behavior of typical conductance/current as a function of 
the difference of two AB fluxes.

The scheme of the paper is as follows. Following the introduction (Section 
$1$), in Section $2$, we describe the model and theoretical formulation 
for the calculation. Section $3$ explores the results, and finally, we 
conclude our study in Section $4$.

\section{Model and synopsis of the theoretical background}

\begin{figure}[ht]
{\centering \resizebox*{6.5cm}{3.5cm}{\includegraphics{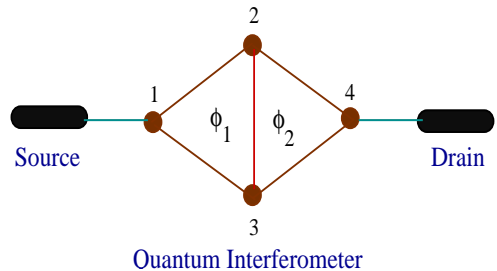}}\par}
\caption{(Color online). Schematic view of a quantum interferometer
with four atomic sites ($N=4$) attached to two semi-infinite 
one-dimensional metallic electrodes.}
\label{ring}
\end{figure}

Let us begin with the model presented in Fig.~\ref{ring}. A quantum
interferometer with four atomic sites ($N=4$, where $N$ gives the total
number of atomic sites in the interferometer) is attached symmetrically 
to two semi-infinite one-dimensional ($1$D) metallic electrodes, namely,
source and drain. The atomic sites $2$ and $3$ of the interferometer are 
directly coupled to each other, and accordingly, two sub-rings, left and 
right, are formed. These two sub-rings are subject to the AB fluxes 
$\phi_1$ and $\phi_2$, respectively.

Considering linear transport regime, conductance $g$ of the interferometer 
can be obtained using the Landauer conductance formula~\cite{land1,land2,
land3,datta,marc},
\begin{equation}
g=\frac{2e^2}{h} T
\label{equ1}
\end{equation}
where, $T$ becomes the transmission probability of an electron across 
the interferometer. It $(T)$ can be expressed in terms of the Green's 
function of the interferometer and its coupling to the side-attached 
electrodes by the relation~\cite{datta,marc},
\begin{equation}
T={\mbox{Tr}}\left[\Gamma_S G_{I}^r \Gamma_D G_{I}^a\right]
\label{equ2}
\end{equation}
where, $G_{I}^r$ and $G_{I}^a$ are respectively the retarded and advanced 
Green's functions of the interferometer including the effects of the 
electrodes. Here $\Gamma_S$ and $\Gamma_D$ describe the coupling of the 
interferometer to the source and drain, respectively. For the complete 
system i.e., the interferometer, source and drain, Green's function 
is defined as,
\begin{equation}
G=\left(E-H\right)^{-1}
\label{equ3}
\end{equation}
where, $E$ is the injecting energy of the source electron. To Evaluate
this Green's function, inversion of an infinite matrix is needed since
the complete system consists of the finite size interferometer and 
two semi-infinite electrodes. However, the entire system can be 
partitioned into sub-matrices corresponding to the individual sub-systems 
and the Green's function for the interferometer can be effectively 
written as,
\begin{equation}
G_{I}=\left(E-H_{I}-\Sigma_S-\Sigma_D\right)^{-1}
\label{equ4}
\end{equation}
where, $H_{I}$ is the Hamiltonian of the interferometer that can be 
expressed within the non-interacting picture like,
\begin{equation}
H_{I} = \sum_i \epsilon_i c_i^{\dagger} c_i + \sum_{<ij>} t_{ij} 
\left(c_i^{\dagger} c_j e^{i\theta_{ij}}+ c_j^{\dagger} c_i 
e^{-i\theta_{ij}}\right)
\label{equ5}
\end{equation}
In this Hamiltonian $\epsilon_i$ gives the on-site energy for the atomic 
site $i$, where $i$ runs from $1$ to $4$, $c_i^{\dagger}$ ($c_i$) is the 
creation (annihilation) operator of an electron at the site $i$ and 
$t_{ij}$ is the hopping integral between the nearest-neighbor sites 
$i$ and $j$. For the sake of simplicity, we assume that the magnitudes of
all hopping integrals ($t_{ij}$) are identical to $t$. The phase factor 
$\theta_{ij}$, associated with the hopping integral $t_{ij}$, comes due to 
the fluxes $\phi_1$ and $\phi_2$ in the two sub-rings. The phase factors 
($\theta_{ij}$) are chosen as, $\theta_{12}=\theta_{23} =\theta_{34}=
\theta_{41}=2\pi\phi/4\phi_0$, $\theta_{24}=2\pi \Delta \phi/2 \phi_0$, 
where $\phi=\phi_1+\phi_2$, $\Delta \phi=\phi_1-\phi_2$ and $\phi_0=ch/e$ 
is the elementary flux-quantum. Accordingly, a minus sign is used for 
the phases when the electron hops in the reverse direction. For the two 
$1$D electrodes, a similar kind of tight-binding Hamiltonian is also used, 
except any phase factor, where the Hamiltonian is parametrized by constant 
on-site potential $\epsilon_0$ and nearest-neighbor hopping integral $t_0$. 
The hopping integral between the source and interferometer is $\tau_S$, 
while it is $\tau_D$ between the interferometer and drain. The parameters 
$\Sigma_S$ and $\Sigma_D$ in Eq.~(\ref{equ4}) represent the self-energies 
due to the coupling of the interferometer to the source and drain, 
respectively, where all the information of this coupling are included into 
these self-energies~\cite{datta}.

The current passing through the interferometer is depicted as a 
single-electron scattering process between the two reservoirs of charge 
carriers. The current $I$ can be computed as a function of the applied 
bias voltage $V$ by the expression~\cite{datta},
\begin{equation}
I(V)=\frac{e}{\pi \hbar}\int \limits_{E_F-eV/2}^{E_F+eV/2} T(E)~ dE
\label{equ8}
\end{equation}
where $E_F$ is the equilibrium Fermi energy. Here we assume that the 
entire voltage is dropped across the interferometer-electrode interfaces, 
and it is examined that under such an assumption the $I$-$V$ characteristics 
do not change their qualitative features. 

All the results in this communication are determined at absolute zero 
temperature, but they should valid even for some finite (low) temperatures, 
since the broadening of the energy levels of the interferometer due to 
its coupling to the electrodes becomes much larger than that of the thermal 
broadening~\cite{datta}. On the other hand, at high temperature limit, all 
these phenomena completely disappear. This is due to the fact that the 
phase coherence length decreases significantly with the rise of temperature 
where the contribution comes mainly from the scattering on phonons, and 
accordingly, the quantum interference effect vanishes. Our unit system
is simplified by choosing $c=e=h=1$. 

\section{Numerical results and discussion}

Before going into the discussion, let us first assign the values of
different parameters those are used for our numerical calculation. The 
on-site energy $\epsilon_i$ of the interferometer is taken as $0$ for 
all the four sites $i$, and the nearest-neighbor hopping strength $t$ 
is set to $3$. On the other hand, for two side attached $1$D electrodes 
the on-site energy ($\epsilon_0$) and nearest-neighbor hopping strength 
($t_0$) are fixed to $0$ and $4$, respectively. The equilibrium Fermi 
energy $E_F$ is set to $0$. 

Throughout the analysis we present the basic features of electron
transport for two distinct regimes of electrode-to-interferometer 
coupling.

\noindent
\underline{Case 1:} Weak-coupling limit
\vskip 0.1cm
\noindent
This limit is set by the criterion $\tau_{S(D)} << t$. In this case,
we choose the values as $\tau_S=\tau_D=0.5$.

\noindent
\underline{Case 2:} Strong-coupling limit
\vskip 0.1cm
\noindent
This limit is described by the condition $\tau_{S(D)} \sim t$. In this
regime we choose the values of hopping strengths as $\tau_S=\tau_D=2.5$.

\subsection{Interferometric geometry with $4$ atomic ($N=4$) sites}

\subsubsection{Conductance-energy characteristics}

\begin{figure}[ht]
{\centering \resizebox*{8cm}{12cm}{\includegraphics{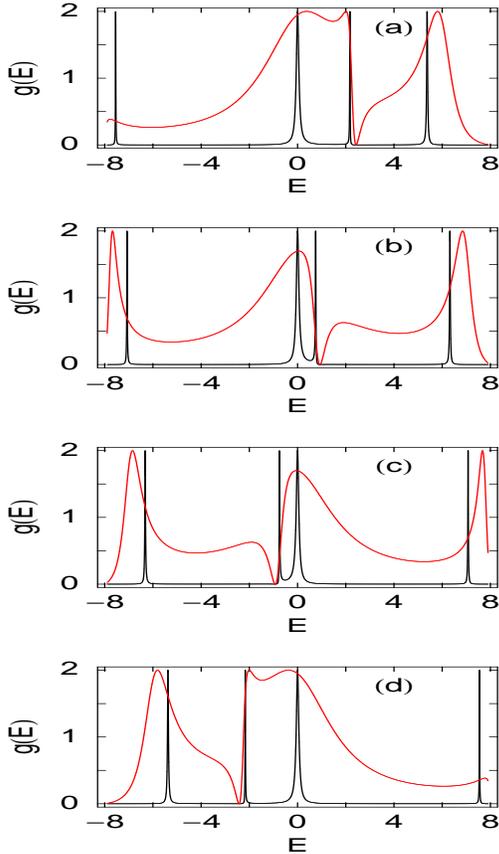}}\par}
\caption{(Color online). $g$-$E$ curves in the weak- (black) and
strong-coupling (red) limits for the interferometer with four
atomic sites ($N=4$) considering $\phi=1$. (a) $\Delta \phi=0.2$, 
(b) $\Delta \phi=0.4$, (c) $\Delta \phi=0.6$ and (d) $\Delta \phi=0.8$.}
\label{cond}
\end{figure}

In Fig.~\ref{cond}, we plot conductance $g$ as a function of the 
injecting electron energy $E$ for the interferometer considering $\phi=1$, 
where (a), (b), (c) and (d) correspond to $\Delta \phi=0.2$, $0.4$, $0.6$ 
and $0.8$, respectively. The black curves represent the results for the 
weak-coupling limit, while the results for the strong-coupling limit are 
shown by the red curves. In the limit of weak-coupling, conductance shows 
fine resonant peaks for some particular energies, while it ($g$) drops to
zero almost for all other energies. At these resonances, conductance 
reaches the value $2$, and therefore, the transmission probability $T$ 
becomes unity, since the relation $g=2T$ is satisfied from the Landauer 
conductance formula (see Eq.~(\ref{equ1}) with $e=h=1$). The transmission 
probability of getting an electron across the interferometer significantly 
depends on the quantum interference of electronic waves passing 
through the different arms of the interferometer, and accordingly, the
probability amplitude becomes strengthened or weakened. Now all the
resonant peaks in the conductance spectra are associated with the 
energy eigenvalues of the interferometer, and thus it is emphasized 
that the conductance spectrum reveals itself the electronic structure of 
the interferometer. The situation becomes quite interesting as long as 
the coupling strength of the interferometer to the electrodes is
increased from the weak regime to the strong one. In the strong-coupling
limit, all the resonances get substantial widths compared to the 
weak-coupling limit. The contribution for the broadening of the resonant 
peaks in this strong-coupling limit appears from the imaginary parts of 
the self-energies $\Sigma_S$ and $\Sigma_D$, respectively~\cite{datta}. 
Hence, by tuning the coupling strength from the weak to strong regime,
electronic transmission across the interferometer can be obtained for
the wider range of energies, while a fine scan in the energy scale is
needed to get the electron conduction across the bridge in the limit
of weak-coupling. These results provide an important signature in the 
study of current-voltage ($I$-$V$) characteristics. Another interesting
feature observed from the conductance spectra is the existence of the
anti-resonant states. The positions of the anti-resonance states can be
clearly noticed from the red curves, compared to the black curves since
the widths of these curves are too small, where they sharply drop to zero 
for the respective energy values associated with the different values of 
$\Delta \phi$ (see Figs.~\ref{cond}(a)-(d)). Such anti-resonant states 
are specific to the interferometric nature of the scattering and do not 
occur in conventional one-dimensional scattering problems of potential 
barriers~\cite{chang,han,he}. A clear investigation shows that the 
positions of the anti-resonances on the energy scale are independent 
of the interferometer-to-electrode coupling strength. Since the width 
of these anti-resonance states are too small, they do not provide any 
significant contribution in the current-voltage ($I$-$V$) characteristics. 

\subsubsection{Typical conductance $g_{typ}$ as a function of 
$\Delta \phi$}

The effect of $\Delta \phi$, the difference between two AB fluxes $\phi_1$
\begin{figure}[ht]
{\centering \resizebox*{8cm}{11cm}{\includegraphics{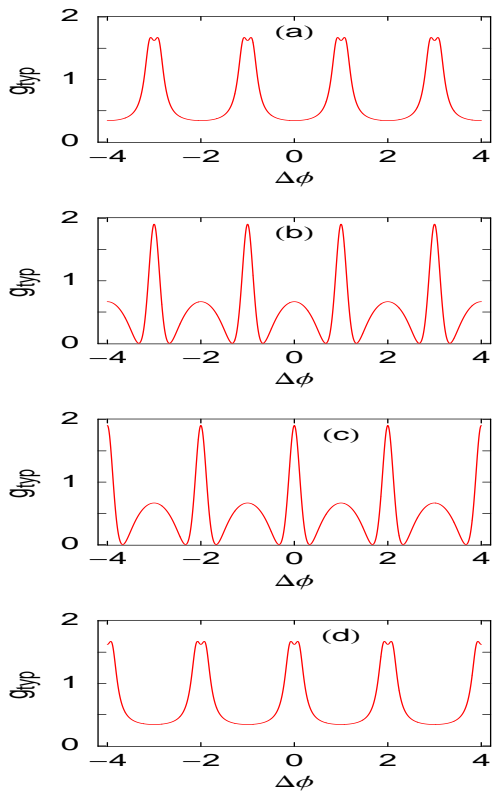}}\par}
\caption{(Color online). $g_{typ}$-$\Delta \phi$ curves in the
strong-coupling limit for the interferometer with four atomic sites
($N=4$), where (a) $\phi=0.2$, (b) $\phi=0.4$, (c) $\phi=0.6$ and 
(d) $\phi=0.8$. The typical conductances are calculated at the energy 
$E=5$.}
\label{cond1}
\end{figure}
and $\phi_2$, on the electron transport through the interferometer is
also an important issue in the present context. To visualize it, in
Fig.~\ref{cond1}, we display the variation of the typical conductance
($g_{typ}$) as a function of $\Delta \phi$ for the interferometer in
the limit of strong-coupling. Figures~\ref{cond1}(a), (b), (c) and (d) 
correspond to the results for $\phi=0.2$, $0.4$, $0.6$ and $0.8$, 
respectively. The typical conductances are calculated for the fixed
energy $E=5$. Very interestingly we observe that, for a fixed value of
$\phi$, typical conductance varies periodically with $\Delta \phi$
showing $2\phi_0$ ($=2$, since $\phi_0=1$ in our chosen unit) 
flux-quantum periodicity, associated with the number of atomic sites ($2$) 
in the vertical line connecting two sub-rings of the quantum interferometer. 
This period doubling behavior is completely different from the traditional 
periodic nature, since in conventional geometries we get simple $\phi_0$ 
flux-quantum periodicity. In the limit of weak-coupling we will also get 
the similar behavior of periodicity ($2\phi_0$) for the typical 
conductance with $\Delta \phi$, and due to the obvious reason we do not 
plot the results for this coupling limit once gain. 

\subsubsection{Current-voltage characteristics}

\begin{figure}[ht]
{\centering \resizebox*{8cm}{10cm}{\includegraphics{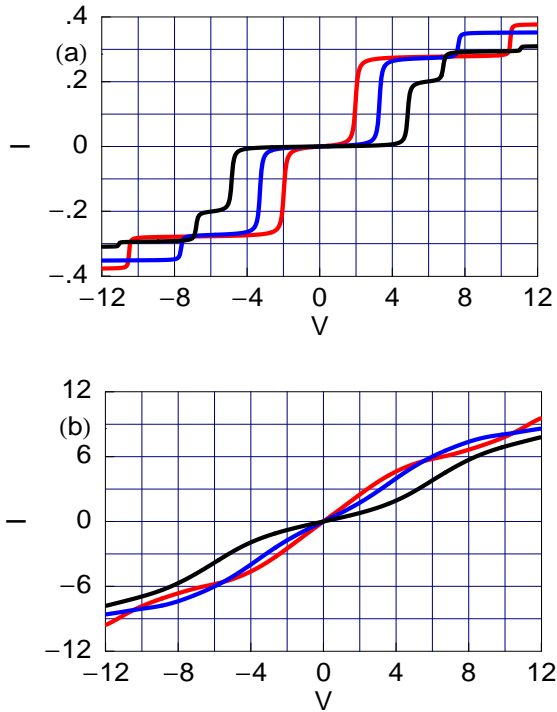}}\par}
\caption{(Color online). $I$-$V$ characteristics of the interferometer
with four atomic sites ($N=4$) for a fixed value of $\phi_1=0.2$, where 
the red, blue and black curves correspond to $\phi_2=0$, $0.1$ and $0.4$,
respectively. (a) Weak-coupling limit and (b) strong-coupling limit.}
\label{current}
\end{figure}

All these features of electron transfer become much more clearly visible
by studying the current-voltage ($I$-$V$) characteristics. The current 
$I$ passing through the interferometer is computed from the integration 
procedure of the transmission function $T$ as prescribed in 
Eq.~(\ref{equ8}) which is not restricted in the linear response regime,
but it is of great significance in determining the shape of the full
current-voltage characteristics. As illustrative examples, in 
Fig.~\ref{current}, we plot the current-voltage characteristics of 
the interferometer for the three different values of $\phi_2$, keeping 
the flux $\phi_1$ in the left sub-ring to a fixed value $0.2$. The red, 
blue and black
curves correspond to $\phi_2=0$, $0.1$ and $0.4$, respectively. In the
limit of weak-coupling (see Fig.~\ref{current}(a)), it is observed that
the current exhibits staircase-like structure with fine steps as a
function of the applied bias voltage $V$. This is due to the existence of
the sharp resonant peaks in the conductance spectrum in this coupling
limit, since the current is computed by the integration method of the
transmission function $T$. With the increase of the bias voltage $V$,
the electrochemical potentials on the electrodes are shifted gradually,
and finally cross one of the quantized energy levels of the interferometer.
Accordingly, a current channel is opened up which provides a jump in the 
$I$-$V$ characteristic curve. The most important feature observed from 
the $I$-$V$ curves for this weak-coupling limit is that, the non-zero
value of the current appears beyond a finite bias voltage, the so-called
threshold voltage $V_{th}$. This is quite analogous to the 
semiconducting nature of a material. Most interestingly, the results 
predict that the threshold bias voltage of electron conduction can be 
controlled very nicely by tuning the AB flux $\phi_2$. The situation 
becomes much different for the strong-coupling case. The results are 
given in Fig.~\ref{current}(b). In this limit, the current varies almost 
continuously with the applied bias voltage and achieves much larger 
amplitude than the weak-coupling case. The reason is that, in the limit 
of strong-coupling all the energy levels get broadened which provide 
larger current in the integration procedure of the transmission function 
$T$. Thus by tuning the strength of the interferometer-to-electrode 
coupling, we can achieve very large, even an order of magnitude, current 
amplitude from the very low one for the same bias voltage $V$, which
provides an important signature in designing nanoelectronic devices.
In contrary to the weak-coupling limit, here the electron starts to
conduct as long as the bias voltage is given i.e., $V_{th}\rightarrow 0$,
which reveals the metallic nature. Thus it can be emphasized that the
interferometer-to-electrode coupling is a key parameter which controls 
the electron transport in a meaningful way. Additionally, the existence 
of the semiconducting or the metallic behavior of the interferometer 
also significantly depends on the AB fluxes $\phi_1$ and $\phi_2$. The 
nature of all these $I$-$V$ curves, presented in Fig.~\ref{current}, will 
be exactly similar if we plot the results for the different values of 
$\phi_1$, keeping $\phi_2$ as a constant. 

\subsubsection{Typical current amplitude $I_{typ}$ as a function of
$\phi_2$}

\begin{figure}[ht]
{\centering \resizebox*{8cm}{10cm}{\includegraphics{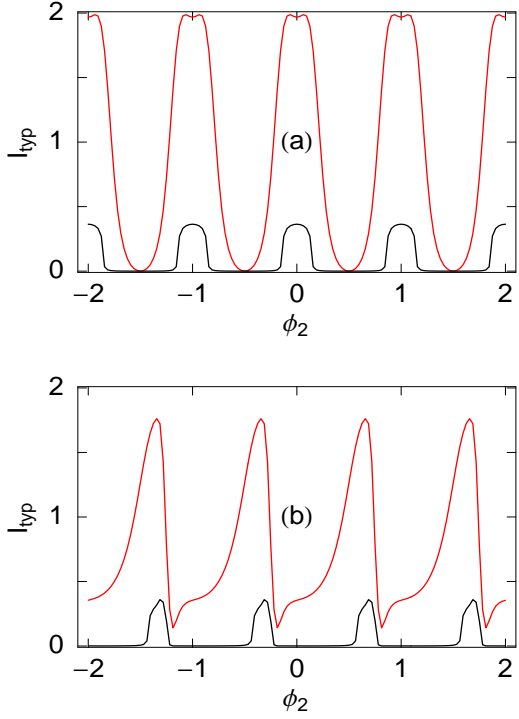}}\par}
\caption{(Color online). $I_{typ}$-$\phi_2$ curves in the weak- (black)
and strong-coupling (red) limits for the interferometer with four
atomic sites ($N=4$), where (a) $\phi_1=0$ and (b) $\phi_1=0.3$. The 
typical current amplitudes are calculated at the bias voltage $V=1.02$.}
\label{typcurr}
\end{figure}

Now, we draw our attention on the variation of the typical current
amplitude with anyone of these two fluxes, when the other one is fixed. 
To explore it, in Fig.~\ref{typcurr}, we show the variation of the typical 
current amplitude ($I_{typ}$) with $\phi_2$, considering $\phi_1$ as a 
constant, where (a) and (b) correspond to $\phi_1=0$ and $0.3$, respectively. 
The black and red lines represent the results for the weak- and 
strong-coupling limits, respectively. The typical current amplitudes are
calculated for the fixed bias voltage $V=1.02$. Both for these two limiting 
cases, the typical current amplitude varies periodically with $\phi_2$, 
exhibiting $\phi_0$ flux-quantum periodicity, as expected. Similar feature 
is also observed for the $I_{typ}$ vs $\phi_1$ curves, when $\phi_2$ becomes 
constant. Here it is also important to note that the variation of $I_{typ}$
with $\Delta \phi$ is quite similar to that as presented in Fig.~\ref{cond1}.
The typical current amplitude varies periodically with $\Delta \phi$
showing $2\phi_0$ flux-quantum periodicity, following the 
$g_{typ}$-$\Delta \phi$ characteristics.

With the above description of electron transport for a $4$-site ($N=4$)
quantum interferometer, now we can extend our discussion for an
interferometer with higher number of atomic sites i.e., $N>4$.

\subsection{Interferometric geometry with $N$ atomic ($N>4$) sites}

\begin{figure}[ht]
{\centering \resizebox*{8cm}{3.75cm}{\includegraphics{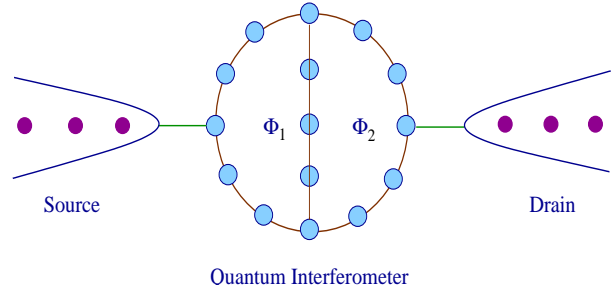}}\par}
\caption{(Color online). Schematic view of a quantum interferometer 
with $15$ atomic sites ($N=15$) attached to two semi-infinite 
one-dimensional metallic electrodes.}
\label{ring1}
\end{figure}

To get an experimentally realizable system, here we 
consider a quantum interferometer with large number of atomic sites 
compared to our presented mathematical model with $4$ atomic sites.
The schematic view of such a quantum interferometer is given in 
Fig.~\ref{ring1}, where we set $N=15$.
The vertical line connecting left and right sub-rings contains $5$
atomic sites, where the individual sub-rings are penetrated by AB
fluxes $\phi_1$ and $\phi_2$, respectively. In this interferometric 
geometry, the phase factors ($\theta_{ij}$'s) are chosen according
\begin{figure}[ht]
{\centering \resizebox*{8cm}{10cm}{\includegraphics{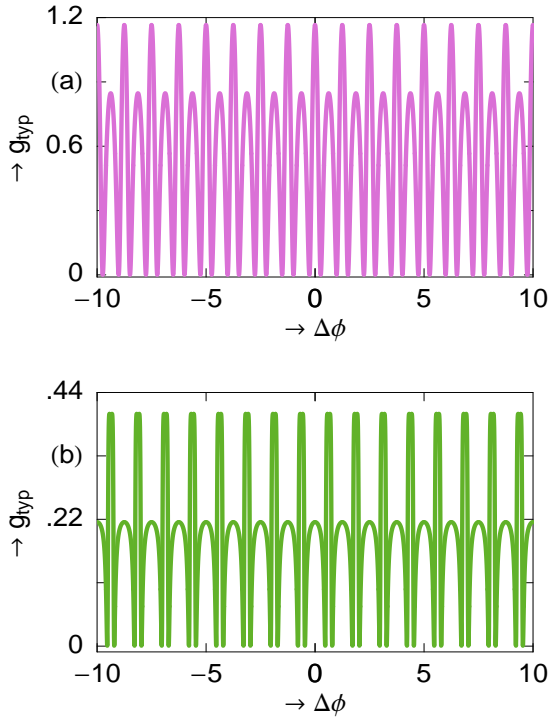}}\par}
\caption{(Color online). $g_{typ}$-$\Delta \phi$ curves in the 
strong-coupling limit for the interferometer with $15$ atomic sites
($N=15$), where (a) $\phi=0.4$ and (b) $\phi=0.8$. The typical 
conductances are calculated at the energy $E=1.5$.}
\label{cond2}
\end{figure}
to our earlier prescription. Along the circumference of the ring
$\theta_{ij}=2\pi \phi/12 \phi_0$ and along the vertical line
$\theta_{ij}=2\pi\Delta \phi/5\phi_0$, where $\phi$ and $\Delta \phi$
correspond to the identical meaning as before.

For this bigger quantum interferometer ($N=15$), exactly 
similar features of conductance-energy and current-voltage characteristics 
are observed as we see in the case of a $4$-site interferometer. Also, 
typical current amplitude $I_{typ}$ shows identical variation with 
$\phi_2$ to our previous study. Only the typical conductance $g_{typ}$ 
varies in a different way as a function of $\Delta \phi$. As illustrative 
examples in Fig.~\ref{cond2} we plot $g_{typ}$-$\Delta \phi$ 
characteristics for the quantum interferometer with $N=15$ in the limit 
of strong-coupling, where (a) and (b) correspond to $\phi=0.4$ and $0.8$, 
respectively. The typical conductances are determined at the energy 
$E=1.5$. From the spectra we notice that for a fixed value of $\phi$,
typical conductance oscillates as a function of $\Delta \phi$ exhibiting
$5 \phi_0$ flux-quantum periodicity. This phenomenon is completely 
different from the traditional periodic nature. Comparing the results
presented in Figs.~\ref{cond1} and \ref{cond2} it is manifested that
the periodicity of $g_{typ}$-$\Delta \phi$ curves depends on the total
number of atomic sites in the vertical line connecting left and right 
sub-rings of a quantum interferometer. Therefore, changing the length
of the vertical line, periodicity can be changed accordingly.

\section{Closing remarks}

To summarize, we have explored electron transport properties through a 
quantum interferometer using the single particle Green's function 
formalism. We have adopted a simple tight-binding framework to illustrate 
the bridge system, where the interferometer is sandwiched between two
electrodes, viz, source and drain. We have done exact numerical calculation 
to study conductance-energy and current-voltage characteristics as functions 
of the interferometer-to-electrode coupling strength, magnetic fluxes 
$\phi_1$ and $\phi_2$ penetrated by left and right sub-rings of the 
interferometer and the difference of these two fluxes. Several key features 
of electron transport have been observed those may be useful in 
manufacturing nanoelectronic devices. The most exotic features are: (i) 
existence of semiconducting or metallic behavior, depending on the 
interferometer-to-electrode coupling strength, (ii) appearance of the 
anti-resonant states and (iii) unconventional periodic behavior of the 
typical conductance/current as a function of the difference of two AB 
fluxes.

Throughout our work, we have addressed the essential 
features of electron transport through a quantum interferometer with 
total number of atomic sites $N=4$. Next, we have extended our discussion
for an interferometer with higher number of atomic sites where we set 
$N=15$ to achieve an experimentally realizable system. In our model 
calculations, these typical numbers ($N=4$ and $15$) are chosen only 
for the sake of simplicity. Though the results presented here change 
numerically with ring size ($N$), but all the basic features remain 
exactly invariant. To be more specific, it is important to note that, 
in real situation experimentally achievable rings have typical diameters 
within the range $0.4$-$0.6$ $\mu$m. In such a small ring, very high 
magnetic fields are required to produce a quantum flux. To overcome 
this situation, Hod {\em et al.} have studied extensively and proposed 
how to construct nanometer scale devices, based on Aharonov-Bohm 
interferometry, those can be operated in moderate magnetic 
fields~\cite{hod1,hod2,hod3,hod4,hod5}.

This is our first step to describe the electron transport in a quantum
interferometer. Here we have made several realistic approximations by
ignoring the effects of electron-electron correlation, electron-phonon 
interaction, disorder, temperature, etc. Over the last few many years 
people have studied a lot to incorporate the effect of electron-electron 
correlation in the study of electron transport, yet no such proper theory 
has been well established. Thus the inclusion of electron-electron 
correlation in the present model is a major challenge to us. The presence 
of electron-phonon interaction in
Aharonov-Bohm interferometers provides phase shifts of the conducting
electrons and due to this dephasing process electron transport through an 
AB interferometer becomes highly sensitive to the AB flux $\phi$ with the 
increase of electron-phonon coupling strength~\cite{hod6}. In the present 
work, we have addressed our results considering the site energies of all 
the atomic sites of the interferometer are identical i.e., we have treated 
the ordered system. But in real case, the presence of impurities will 
affect the electronic structure and hence the transport properties.
The effect of the temperature has already been pointed out earlier, and, 
it has been examined that the presented results will not change 
significantly even at finite temperature, since the broadening of the 
energy levels of the interferometer due to its coupling to the electrodes 
will be much larger than that of the thermal broadening~\cite{datta}.
At the end, we would like to mention that we need further study in such 
systems by incorporating all these effects.

The importance of this article is mainly concerned with (i) the simplicity 
of the geometry and (ii) the smallness of the size, and our exact analysis
may be utilized to study electron transport in Aharonov-Bohm geometries.

\end{document}